 \documentclass[]{article}    
 \usepackage{spie}            

\input{psfig}

\title{GAIA:  ORIGIN AND EVOLUTION OF THE MILKY WAY}

\author{
Gerry Gilmore\supit{1}, Michael Perryman\supit{2}, 
L Lindegren\supit{3}; \\  F Favata\supit{2}; E Hoeg\supit{4}; M
Lattanzi\supit{5}; 
X Luri\supit{6}; \\ F Mignard\supit{7}; S Roeser\supit{8};P.T.
deZeeuw\supit{9}
\skiplinehalf 
\supit{1}Institute of Astronomy, Madingley Road, Cambridge, CB3 0HA,
UK  \authorinfo{E-mail contacts: gil@ast.cam.ac.uk; mperryma@astro.estec.esa.nl}
\skiplinehalf 
\supit{2}Astrophysics Division, ESTEC, Noordwijk 2200AG, The
Netherlands 
\skiplinehalf 
\supit{3}Lund Observatory, Sweden; 
\supit{4}Copenhagen University Observatory, Denmark; \\
\supit{5}Torino Observatory, Italy;
\supit{6}Universitat de Barcelona, Spain; \\
\supit{7}CERGA, Grasse, France;
\supit{8}ARI Heidelberg, Germany;\\
\supit{9}Sterrewacht Leiden, P.O. Box 9513, 2300 RA, Netherlands
}

\begin{document} 

\maketitle 


\begin{abstract}

GAIA is a short-listed candidate for the ESA Cornerstone mission C5,
meeting the ESA Survey Committee requirement for an observatory
mission, dedicated to astrometry, providing 10 micro-arcsecond
accuracy at 15th magnitude.  The GAIA mission concept follows the
dramatic success of the ESA HIPPARCOS mission, utilising a
continuously scanning spacecraft, accurately measuring 1-D coordinates
along great circles, in (at least) two simultaneous fields of view,
separated by a known angle. These 1-D relative coordinates are later
converted to the five astrometric parameters of position and motions
in a global analysis.  GAIA will provide precise astrometry and
multi-colour photometry for all the one billion stars, quasars, and
compact galaxies to I=20 on the sky.  GAIA will additionally provide
the sixth phase-space parameter, radial velocity, from a slitless
spectroscopic survey of most stars brighter than about magnitude 17.
The technical challenges are considerable, but achievable.  The
scientific returns are spectacular, with greatest impact in the study
of stellar populations and dynamical structure of the galaxies of our
Local Group, and in providing the first complete census of the stars
and massive planets in the Solar neighbourhood. GAIA will
revolutionise our knowledge of the origin and evolution of our Milky
Way Galaxy, and of the distribution of planetary systems around other
stars.

\end{abstract}




\section{GAIA: THE ESA CONTEXT}

The European Space Agency supports a series of major missions, known
as {\tt cornerstones\footnote{ see http://www.estec.esa.nl/spdwww/ }},
complemented by a larger number of smaller projects. The special
feature of cornerstone missions is that they serve a large community
in a discipline of long-term importance.  Forthcoming cornerstone
missions include XMM (C2, launch 1999), Rosetta (C3, launch 2003), and
FIRST (C4, launch 2006). The provisional timetable for Cornerstone C5
involves final selection in the next two years, for scheduled launch
in 2009.

In 1992 a Survey Commitee was assembled at the request of the ESA
Council to establish scientific priorities for its long-term
scientific programme, in the post-Horizon 2000 era.  Following on from
the success of the first space based dedicated astrometric project,
HIPPARCOS, the committee recommended ``that ESA initiate a
Cornerstone-level programme in interferometry for use as an
observatory open to the wide community. The first aim [of which] is to
perform astrometric observations at the 10 microarcsec level."

Absolute astrometry at this level of accuracy for large samples of
stars constitutes a very powerful scientific tool, allowing
investigation of the structure, contents, and dynamics of the Milky
Way, through determination of the distances and motions, and detection
of possible planetary or brown dwarf companions, of very large samples
of stars, as well as study of the dynamics of nearby external
galaxies. To this end the GAIA mission is under study to address the scientific
requirements and technological challenges implied by such a
mission. Two industrial studies are underway, by Alenia (It) and Matra
Marconi Space (Fr), to define complementary approaches to the payload
design. The Alenia study considers a Fizeau interferometer
design, with precision metrology and mechanical control; the Matra
Marconi study has concentrated on a payload in which the requisite
extreme stability is provided by a stable passive monolithic
structure.

It is useful to note that the only currently approved space
astrometric mission is the NASA project SIM. GAIA and SIM are
remarkably complementary in both design and scientific goals, though
should deliver rather similar limiting astrometric accuracy. SIM is
primarily a technology precursor mission for the Origins programme,
and so is necessarily based on long-baseline Michelson
interferometers. Any such instrument must operate in pointed mode, on
a pre-selected target list. SIM will provide superbly accurate
astrometry for of order $10^4$ stars over a five-year mission life.
It is thus ideal for rare but intrinsically interesting classes of
stars.  GAIA is a scanning satellite, following the proven HIPPARCOS
model of operation, with fewer constraints on its mode of
implementation.  Thus, GAIA may adopt a design which reduces
technological complexity. GAIA is optimised for very large samples, up
to $10^{9}$ stars. It is thus ideal for providing an accurate global
picture of the Milky Way based on reliable statistical samples, for
the discovery of new classes of rare object, and for searches for
planetary systems.

\section{THE ORIGIN AND EVOLUTION OF GALAXIES: \\ AN OBSERVATIONAL CHALLENGE}

\label{sect:intro}  

   \begin{figure}

   \vspace{10cm}  
   { \label{fig:example1}	  

FIGURE 1: An illustrative summary of the primary GAIA scientific goals,
superimposed on the Lund map of the Milky Way and Local Group
galaxies.}

   \end{figure}

Understanding the origin, past evolution, present-day structure,
stellar and planetary contents, and long-term future, of the part of
the Universe in which we live ranks among the great intellectual
challenges facing modern science. The current astrophysical paradigm
is that galaxies like the Milky Way grew, primarily under the
influence of gravity, from small-amplitude fluctuations in the very
early Universe. The growth of these fluctuations, and all later
structures, is dominated by the nature and amount of the apparently
ubiquitous but as yet unidentified dark matter. As the early
fluctuations in the dark matter and primordial hydrogen and helium
grow into collapsing and cooling clouds, stars will form, the most
massive of which rapidly evolve, create new chemical elements, and
disperse them through supernovae. The lower mass stars, and their
planetary systems, survive to the present day. The star-forming gas
clouds and their dark matter halos themselves aggregate into larger
and larger structures, containing more and more of the chemical
elements, eventually forming the dramatic diversity of types of Galaxy
which exist today, including the Milky Way Galaxy and its several
dwarf satellite companions, and the several other neighbour galaxies
which comprise our Local Group.

This broad-brush summary, while plausible, remains to be tested in
detail, and depends sensitively on several key processes -- the
behaviour of dark matter, the nature of star formation, the incidence
of planetary systems -- which remain entirely unknown. Is progress
possible? Perhaps surprisingly, understanding the complex origin of
the Milky Way is readily amenable to progress by direct observation.
This process of Galacto-archaeology is in principle feasible as a
direct observation because typical low-mass stars live for longer than
the present age of the Universe, and preserve in their atmospheres an
(almost) unmodified record of the chemical elements from which they
were formed.  Thus, the numbers and chemical element abundances of
long-lived stars provide a direct fossil record of the creation
history of the chemical elements, and of the rate at which stars have
formed since the Big Bang. In addition, their orbits encode the
crucial complementary information needed to understand galaxy
formation and evolution: analysis of the spatial distributions of
stars and their orbital energies traces the otherwise unobservable
spatial distribution of dark matter; departures from smoothness in the
distribution of stars in velocity and coordinate (phase) space retain
a memory of the rate and time at which large galaxies have grown by
devouring their smaller companions.  For stars near the Sun, planetary
companions induce observable wobbles on the orbit through the Milky
Way of their parent sun, allowing their detection from precise
position and kinematic data.

Are such observations possible in practice? Yes indeed. The required
observational products are three spatial coordinates, two providing a direction,
the third a distance down the line of sight, and three orthogonal velocity
coordinates, two in the plane of the sky and the third radial component.
Ideally, for at least a large subset of the targets, one additionally
desires a measure of the abundance of the chemical elements
-- obtainable from analysis of the stellar spectrum -- and a measure of
the age -- derivable in special cases from analysis of photometry if the
distance is known. While in principle one might obtain such
information for all $10^{11}$ stars in a galaxy like the Milky Way, 
in practice, fortunately, only a  representative sub-set need be observed.
A critical requirement of course, without which the value of
any data set is substantially degraded, is very accurate knowledge of
the selection function which defines the sub-set of about $10^9$ stars
that will be observed.
This implies accurate photometry to faint limiting magnitudes, and
high spatial resolution imaging, since much of the sky is crowded, or occupied
by compact background galaxies or interstellar emission.

Thus the requirements for a direct observational determination of the origin,
content, structure and evolution of the Milky Way Galaxy become well
defined: photometry, spectra and astrometry. \\

\noindent{i)} A photometric survey, to provide accurate definition of the
subset of all stars which are to be studied further, and some
complementary age information; \\

\noindent{ii)} A radial velocity survey, to provide one of the three
components of the velocity vector, and complementary chemical
abundance information from the requisite spectra; \\

\noindent{iii)} Three position coordinates, two of which are angles, the
third requiring a parallax distance; \\

\noindent{iv)} The two components of the transverse velocity vector, which
appear as proper motions on the plane of the sky.

\medskip
One may immediately define the required performance to meet these
science goals. In order to probe the whole Milky Way and its
satellites using reasonably understood albeit intrinsically luminous
stellar tracers, the proper motion survey must reach fainter than about I=18, while
the radial velocity survey must reach an accuracy of a few km/s to
apparent magnitudes of about I=17 ({\sl cf} Gilmore \& Hoeg 1995; Gilmore \&
Perryman 1997 for details).  A representative space velocity for an outer disk
star, or a star in a representative stellar cluster, is of order
100$\mu$as/yr, implying a required precision of order 10$\mu$as.
Detection of solar system-like planetary systems, if they exist,
around the nearest 100,000 stars requires similar astrometric
precision, of about 10$\mu$as.

Provision of precise astrometric information for bright stars has been
proven possible by the success of the ESA HIPPARCOS satellite mission.
GAIA is being developed to build upon the HIPPARCOS success, extending
performance by some two orders of magnitude in precision, some four
orders of magnitude in sensitivity at the sample completeness limit,
and four orders of magnitude in sample size. With these gains Galactic
archaeology will become a real observational science, and we will have
our first accurate census of the stars -- and planetary systems -- in
our local neighbourhood.

Figure~2 illustrates
the limitations in our present knowledge of even our immediate
Galactic neighbourhood, and the gains which await from GAIA.

   \begin{figure}

   \vspace{10cm}  


   { \label{fig:example2}	  

FIGURE 2: The three panels illustrate a slice of sky, centred on the
Sun, $ \rm 90^{\circ} \times 10^{\circ}$ on a side, containing the
Hyades open star cluster. Each point represents one star, with the
uncertainty in its line of sight distance illustrated where it exceeds
the size of the point.  The left hand panel shows the current state of
the art of ground-based measurements; the centre panel the HIPPARCOS
state of the art; the right hand panel a few percent of the sources in
the sky as it will be mapped by GAIA.  } \end{figure}


\section{ ASTROPHYSICS WITH GAIA}

A detailed photometric and kinematic survey of a billion stars is a
major undertaking. Would not a million do?  One should therefore start
by asking just why large samples are needed. The table below lists the
most obvious subset of the GAIA scientific case, while further details
and examples may be found in Gilmore \& Hoeg (1995); Gilmore \&
Perryman (1997); and Perryman, Lindgren \& Turon (1997). The general
goals of the GAIA science program are summarised in Figure~1 above. To
further illustrate the need for large sample statistical astrophysics
we present three brief outlines of primary GAIA science: the origin of
the Milky Way Galaxy, the distribution of dark matter, and the number
of planetary systems near the Sun.

\begin{table} [h]   
\caption{
A brief summary of the GAIA scientific case }

\begin{center}       
\begin{tabular}{|l|l|} 
\hline
\rule[-1ex]{0pt}{3.5ex} THE HISTORY OF OUR GALAXY  & test hierarchical
formation theories  \\
\hline
\rule[-1ex]{0pt}{3.5ex}   &  inner bulge/bar dynamics \\
\hline
\rule[-1ex]{0pt}{3.5ex}   &  disk-halo dynamical interactions \\
\hline
\rule[-1ex]{0pt}{3.5ex}  & continuing dynamical evolution  \\
\hline
\rule[-1ex]{0pt}{3.5ex}  & what is the warp  \\
\hline
\rule[-1ex]{0pt}{3.5ex}  & star cluster disruption  \\
\hline
\rule[-1ex]{0pt}{3.5ex}  &  weigh spiral structure  \\
\hline
\rule[-1ex]{0pt}{3.5ex}  &  star formation history  \\
\hline
\rule[-1ex]{0pt}{3.5ex}  &  chemical evolution  \\
\hline
\rule[-1ex]{0pt}{3.5ex}  &  link to high redshifts \\
\hline
\rule[-1ex]{0pt}{3.5ex} STELLAR EVOLUTION &  large sample to
detect rapid evolutionary phases  \\
\hline
\rule[-1ex]{0pt}{3.5ex}  &  quantify pre-main sequence evolution \\
\hline
\rule[-1ex]{0pt}{3.5ex}  &  complete census of the local neighbourhood \\
\hline
\rule[-1ex]{0pt}{3.5ex} STELLAR FORMATION & dynamics of star forming regions    \\
\hline
\rule[-1ex]{0pt}{3.5ex}  &  luminosity function for pre-Main Sequence stars \\
\hline
\rule[-1ex]{0pt}{3.5ex} BROWN DWARFS &  census of brown
dwarfs in binaries \\
\hline
\rule[-1ex]{0pt}{3.5ex} PLANETARY SYSTEMS &  complete census around
$\sim 3.10^5$ stars\\
\hline
\rule[-1ex]{0pt}{3.5ex} THE LOCAL GROUP &  rotational parallaxes for
Local Group galaxies \\
\hline
\rule[-1ex]{0pt}{3.5ex}  &  kinematic separation of stellar populations \\
\hline
\rule[-1ex]{0pt}{3.5ex}  &  galaxy orbits to give cosmological history \\
\hline
\rule[-1ex]{0pt}{3.5ex} BEYOND THE LOCAL GROUP &  parallax calibration
of distance scale \\ 
\hline
\rule[-1ex]{0pt}{3.5ex}  &  zero proper motion QSO survey  \\
\hline
\rule[-1ex]{0pt}{3.5ex}  & photometry of $10^8$ galaxies  \\
\hline
\rule[-1ex]{0pt}{3.5ex}  THE NATURE OF  MATTER & Galactic
rotation curve  \\ 
\hline
\rule[-1ex]{0pt}{3.5ex}   &   disk mass profile from $\sigma_z(R)$ \\
\hline
\rule[-1ex]{0pt}{3.5ex}   &  internal dynamics of Local Group dwarfs \\
\hline
\rule[-1ex]{0pt}{3.5ex}  FUNDAMENTAL PHYSICS &  Determine the
space-curvature parameter $\gamma$ to $10^{-6}$.   \\
\hline
\rule[-1ex]{0pt}{3.5ex}  REFERENCE FRAMES &  Define the local metric \\
\hline
\rule[-1ex]{0pt}{3.5ex} SERENDIPITY &  the first all-sky phase-space map..... \\
\hline

\end{tabular}
\end{center}
\end{table}

\subsection{GALACTIC ASTROPHYSICS WITH GAIA}

The kinematic and metallicity distribution functions of complete
samples of long-lived stars have long been recognised as providing
unique constraints on the early stages of  evolution of the chemical elements
in the Galaxy.  The main sequence lifetime of low-mass dwarf stars is greater
than the age of the Galaxy; the chemical-abundance distribution
function of such stars provides an integrated record of the
chemical-enrichment history without the need for model-dependent
corrections for dead stars.  Pioneering studies focussed on the only
reasonably-complete sample available, which is that for solar-like
stars in the immediate solar neighborhood; in effect the brightest
stars within about 30pc of the Sun.  These samples have been
sufficiently small that reliable study of those stellar populations
which are not common in the solar neighborhood has necessarily been
difficult. This is  a serious restriction, as such stars
might in principle be a major contributor to the stellar population in
a valid, representative volume of the Galaxy.  For example, stars of
the Galactic inner disk and bulge, and outer halo, are barely
represented, if at all. In addition, intrinsically-rare stellar
populations and short-lived phases of stellar evolution are missed
entirely, even though they may be crucial for understanding important
physics, or significant events in Galactic history.

The observational situation has been improved recently in three ways:
by collection and analysis of spectroscopic data for all-sky samples
of stars extending to somewhat greater, but still essentially local,
distances; by deeper pencil-beam surveys, to isolate {\sl in situ}
samples of old disk, thick disk and halo stars; and most dramatically,
by the HIPPARCOS mission which has quantified local kinematics. The
combination of the large ($\sim10^5$ stars) but local samples with the
smaller but distant samples ($\sim10^4$ stars) has allowed the
deconvolution, to first order, of the abundance distribution
functions, and the mean velocity dispersions, of the dominant Galactic
populations. While our understanding of Galactic structure and
evolution has advanced considerably of late, extension of these
analyses has become limited by the intrinsic breadth and overlap of
the population distribution functions and by the small size of the
available {\sl in situ} samples.

The theoretical situation has also become more specific.  Though the
many dynamical, structural and chemical evolution questions one poses
concerning galactic evolution may seem well-defined and relatively
distinct, it is now clear that the answers are intimately
interrelated. For instance, galaxies probably accrete their
neighbours, so that the place of origin of a star may be far from its
present location; dynamical instabilities in disks result in the
mixing through phase space of stellar populations,  blurring
the relation between a star's present location and its birthplace. Bar
instabilities are also likely to cause significant gas transport, and
may drive star bursts and possibly nuclear non-thermal phenomena.
Major mergers may thicken disks. Bulges may be accreted, or created
during mergers.

Modern models of Galaxy formation make fairly specific predictions
concerning each of these possibilities.  For example, fashionable Cold
Dark Matter models, which contain aspects of both the monolithic
(`ELS') and the multi-fragment (`Searle-Zinn') pictures often
discussed in chemical evolution models, `predict' growth of the Galaxy
about many small cores, which should contain the oldest stars. After
several of these merge, defining what is to become the eventual Milky
Way Galaxy, the new central regions will retain the oldest
stars. Subsequent accretion of small `galaxies' forms the outer halo
and the disks, while late accretion will continue to affect the
kinematic structure of both the outer halo and the thin disk.  The
normal star formation in the disk produces clusters and
associations. The latter are unbound at birth, and the former lose
stars due to internal dynamical evolution and the Galactic tidal
field, resulting in a multitude of moving groups o fstars with a
common origin.  Considerable phase-space substructure should thus be
detectable almost anywhere in the Galaxy.

Dissipational models for thick-disk formation predict observable
spatial gradients in the distribution of the chemical elements, 
and similar scale lengths for the thick and thin Galactic
disks.  Specific column-integral element abundance distributions can be
calculated (numerically) for some of these models and compared to
observations.  Satellite merger models for thick disk formation
require the stars from the satellite to be detectable, as a tail in
the thick disk distribution functions below [Fe/H]=$-1$.  `Continuum'
models of thick disk formation from the thin disk require that an
accurately defined joint distribution function over chemical abundance
and kinematics for the oldest stars be smooth and continuous.
Alternative models, such as the discrete merger model, can then be
distinguished by their prediction that the distibutions overlap in
abundance, and perhaps velocity dispersion, but not in angular
momentum (Gilmore, Wyse \& Kuijken 1989).  Most detailed models
make specific predictions concerning the abundance distribution
function in a cylinder, through the Galactic disk - the `G-dwarf
problem' -- which remains widely studied, and a valuable diagnostic of
early accretion and gas flows in the disk.  Extension to a more
representative volume is necessary for Galactic-scale analyses.
Some further details are presented in Gilmore \& Wyse (1986).

That is, quantitative study of the essential physics of galaxy
evolution requires that one must study the distributions over
chemistry, kinematics and spatial structure of a large and
representative sample of stars. Such a sample can be provided only by
accurate measurement covering a substantial fraction of all of the
phase space potentially available to stellar populations. In coordinate terms, one
must measure the stars where they are, across most of the volume of
the Galaxy. In number terms, one must measure large enough samples in
each place that the local properties, and their gradients, can be
known reliably. Since one does not know, prior to measurement, where a
star is, one must measure very large numbers of stars in each
direction to be confident of one's conclusions.  The ambitious goal of
understanding the Milky Way requires an ambitious experiment.

\subsection{DARK MATTER STUDIES WITH GAIA}

The nature of the dark matter which apparently dominates the evolution
of structure in the Universe, and the dynamics of every galaxy, is one
of the biggest questions in contemporary physics. Extensive analyses
suggest that the smallest length scale on which dark matter is
gravitationally dominant is that of the dwarf satellite galaxies of
the Milky Way. Thus these galaxies provide the best possible testbed
to determine the density, velocity dispersion, and 3-D distribution
of the dark matter. Such studies are complemented, and linked, to those
on cosmological scales by measurements on larger length scales, which
are attainable by mapping the dynamics of the Milky Way and the Local
Group.

The technique to achieve this requires two approaches: detailed dynamical
mapping of the dwarf satellites of the Milky Way, especially
Sagittarius, the nearest, and the Large Magellanic Cloud, the largest;
complemented by detailed determinations of
the rotation curve and disk mass of the Galaxy as a function of
radius.  It has been shown that extant data, with at most a few
hundred accurate radial velocities and little or no useful proper
motion data, is inherently unable to achieve this dynamical
mapping. The problem is degeneracy between anisotropic stellar orbits,
supporting the galaxy by stress, and gravitational potential
gradients, supporting the galaxy by pressure.  Extant data cannot
distinguish robustly between dark matter distributions which are very
centrally concentrated, are moderately extended isothermal, or are
very extended. An ability to distinguish between these
different allowed dark matter distributions is clearly a key step for
progress.

This degeneracy can be broken with sufficiently accurate and
sufficiently extensive proper motion and radial velocity data. Current
simulations suggest the need for some thousands of stars per satellite
galaxy, and some millions throughout the Galactic disk, for this
experiment.

\subsection{DETECTION OF PLANETARY SYSTEMS}

In order to understand the numbers of planetary systems in the Milky
Way we need to know at minimum how they are distributed with different
parent star properties, and what range of orbital sizes they have. Any
such understanding requires careful analysis of a large and complete
sample of stars to see if they are parents of planetary systems, or
are isolated. Astrometry provides the ideal tool for this experiment.

The reflex motion of the Sun induced by Jupiter's orbital motion is
equivalent to 500$\mu$as at 10pc; that of the 3-year period planet on
its primary 47UMa is 362$\mu$as. These values illustrate the utility
of accurate astrometry with a time base of several years for planetary
detection. Careful simulations show that GAIA has a detection
efficiency for a Sun-Jupiter system which exceeds 50\% at 150pc,
covering $\sim 3.10^5$ stellar systems, and is still useful at 200pc
(Lattanzi, Spagna, Sozzetti \& Casertano, 1997).

This sensitivity is complemented by the other special feature of the
GAIA mission: its unique capability to identify the complete sample of
stars within the few hundred parsecs centred on the Sun. Thus GAIA
will be able to survey the nearest 300,000 stars for planets, because
it will identify those 300,000 stars. This will allow accurate
determination of the frequency of planetary systems as a function of
the type of parent star, as well as of planetary mass and separation.
Such a data set will of course provide, at the same time, a superb
determination of the number of brown dwarfs -- objects between
planetary and stellar mass -- which are invisible binary companions to luminous
stars. This distribution is a key parameter in understanding accretion
rates and angular momentum transport in star formation, but is at
present unknown.

\subsection{FUNDAMENTAL PHYSICS and REFERENCE FRAMES}

Very precise determination of the spatial reference frame is a
critical requirement in many fields: in astrophysics from
inter-planetary navigation to relative location of observations at
different wavelengths.  GAIA is both precise in individual
measurements, and dense in numbers of observations. Thus it will
enhance both the local density and the global integrity of the
reference frame. GAIA will observe an all-sky grid of cosmologically
distant point sources which are observable at (almost) all wavelengths
from radio to X-ray -- the quasars -- and use these to improve both
the zero-point global determination of the astronomical reference and
its local implementation in any specific line of sight.

Any astrometric measurement at micro-arcsec scales is dominated by the
systematic distortions of the local metric generated by local masses,
especially the Sun and major planets.  Light bending by the Sun is
4000$\mu$as even at the ecliptic pole, 90$\rm ^{\circ}$ from the Sun;
by the Earth is 40$\mu$as at its limb, 17000$\mu$as at Jupiter's
limb. Precise determination of these systematic effects allows
determination of both the time-space and space-space components of the
metric tensor, at a level which is of interest for some viable
scalar-tensor theories of gravitation.  The combined effects are related to the
well-known geodetic precession and gravito-magnetic (Lense-Thirring)
effects, often quantified through the parameterised post-Newtonian
(PPN) parameter $\gamma$. Light deflection uniquely is sensitive to
only the space-space part of the metric, allowing sensitive
determination of its amplitude ({\sl cf} de Felice, Lattanzi,
Vecchiato \& Bernaca 198).
$\gamma$ is currently known to be equal to
the General Relativistic prediction to one part in $10^{-3}$.  GAIA
will determine $\gamma$ to one part in $10^{-6}$.

\section {THE GAIA SPACECRAFT}

Achievement of the ambitious scientific goals noted above imposes
demanding technical performance specifications on a spacecraft. The
most obvious requirement is stability during the time it takes  to
collect the large number of photons needed for an accurate location
of an image. To appreciate this it is helpful to recall the principles
of global astrometry, as implemented by the HIPPARCOS mission.

   \begin{figure}

   \vspace{10cm}  
%


   { \label{fig:example3}	  

FIGURE 3: An exploded view of the GAIA spacecraft, superimposed on an
HST image of part of the Large Magellanic Cloud, one of GAIA's prime
scientific targets. The spacecraft support module is at the left hand
side, separated by a large sunshield from the payload module. The
possibly unnecessary 400N motor is at the extreme left, at the Sun and
Earth pointing end. Next is shown the ring of phased-array antennae,
supporting the high telemetry requirement. Solar power arrays and the
sunshield are shown, followed by the science payload. This is a set of
three telescopes on a single support structure, which also supports
the focal planes and electronics. The whole spacecraft spins as a
single structure about the horizontal axis in this view.  This
arrangement provides the essential thermal and mechanical stability
for GAIA.  } \end{figure}

\subsection{GLOBAL ASTROMETRY}

The fundamental measurement in any astronomical astrometric system is
of the angle between two or more stars. Global astrometry extends this
concept by determining all the angles between all the stars in a large
sample covering the whole celestial sphere. It is then possible to use
this multiply redundant measurement set to solve for a reference grid
with rigidly orthogonal coordinates over the whole sky. One
additionally requires a set of zero-point reference objects, to
determine the origin and rotation of this global rigid grid. 
The crucial requirement is to have a spinning spacecraft whose orbit
precesses its lines of sight over the whole celestial sphere.

To obtain their sets of relative positions, HIPPARCOS used, and GAIA
will use, two lines of sight at a rigidly fixed angle (basic angle)
perpendicular to the spacecraft spin axis. This allows an
instantaneous measurement of the differential positions between all
the sources (stars, galaxies, quasars, asteroids, planets...) in the
two fields. A little later the spacecraft has rotated slightly,
leaving some sources behind, and adding new sources for additional
relative position measurements.  As the spacecraft spins (GAIA will
have a spin period of about 3 hours) the scanning path on the sky of
the two lines of sight will allow measurement over one complete spin
period of relative positions for all objects around the great
circle. An observation proceeds by clocking the focal plane CCD
detector in TDI mode at the spacecraft spin rate, integrating charge
during the transit time of an image
across the focal plane. As a star falls off the trailing edge of the
focal plane, the area containing its integrated image is readout, the
time noted, and both transmitted to ground for later centroiding and
analysis.

Precession moves the scanned circle around the sky, allowing whole sky
coverage. Repeating this some 100 times over five years gives one
sufficient information to derive not only the positions of everything
observed relative to everything else observed, but also a sufficiently
long time baseline that one may additionally determine the transverse
(proper) motions of everything relative to everything else, as well as
the apparent motion of relatively nearby sources due to the annual
motion of the  Earth around the Sun, or parallax, which directly provides the
stellar distances.  Solving this very large data set for all these
parameters is a well-defined though not straightforward task. It is
then necessary to set zero points relative to non-moving objects: in
the case of a satellite with high accuracy, the only known objects
which define a suitable (Machian) reference frame are those at
cosmological distances, among which quasars are the ideal. GAIA will
use more than $10^5$ quasars to define its reference system.

\subsection{OPTICAL DESIGN and FOCAL PLANE} 

The primary design constraint for GAIA is that it be able to be
launched in an Ariane5 shroud. This provides an upper limit on the
optical baseline of any monolithic payload of about 2.5m. Given launch
costs, restriction to a shared launch with a second independant
payload is highly desirable, which provides a size and mass limit.

Both Fizeau and monolithic payloads are under evaluation for
GAIA. At the time of writing consideration of the monolithic payload
is more advanced, so it is that option which is summarised here. Analysis
of the Fizeau interferometer system made clear that precise
(nm) mechanical control is most easily provided by combining the
separate apertures onto a single monolithic mirror, and freeing that
mirror from mechanical and thermal disturbances. It is a relatively
small step from there to use of the full, albeit very rectangular,
primary mirror. An extra advantage of such a strategy is to minimise
the number of independently controlled sub-systems, regardless of
their required accuracy, significantly assisting system simplicity and
reliability.

The current GAIA design, illustrated in the exploded view above, has
three telescopes, each a three-mirror system, mounted around a
carrying strut. Two telescopes are identical, and provide the two
optical systems necessary to allow global astrometry. The third
is fitted with a slitless spectrograph, and provides the spectra for
radial velocity and chemical element abundance determination.

The two astrometric telescopes have a primary mirror of 1.7 by 0.7m,
with the spectroscopic telescope being somewhat smaller. The entire
structure is SiC, to maximise rigidity and thermal uniformity while
minimising weight.

The astrometric telescopes are currently assumed to have a focal
length near 50m, allowing use of high efficiency currently available
CCDs as detectors. The focal planes cover somewhat more than 0.5sq deg
with acceptable image quality, requiring some 250 independent
CCDs for each astrometric telescope. 
The outer parts of the focal plane provide sufficient image
quality for on-board image detection. This is a particularly important
advantage over HIPPARCOS, where pre-launch selection of targets was
necessary. As emphasisied in the scientific case above, pre-selection
of targets significantly restricts the scientific potential of the mission.
In fact, on present performance expectations, the single factor
limiting the number of sources which can be measured by GAIA is the
telemetry rate in providing useful data to ground.

The radial velocity telescope is similarly a three-mirror system,
differing from the primary astrometric systems only in shorter focal
length, and in addition of a slitless spectroscopy capability. Spectra
are integrated during a focal-plane transit, readout, and transmitted
down for radial velocity and chemical element abundance analysis.

\subsection{ORBIT and OPERATIONS} 

An obvious advantage for any precise system is location in a benign
environment, though such environments invariably possess some
disadvantages.  For GAIA the benign environment involves location at
the outer Lagrangian point L2, some 1.5million km from Earth, in a
maximally stable thermal environment. The cost is reduced
telecommunications capability, and possibly increased on-board motor
requirements. This latter requirement, for an on-board 400N capability
to reach L2 after launch into geostationary orbit and release, may
become unnecessary, if the expected restart mode of Ariane 5 is
implemented. Direct injection of GAIA into L2 by the Ariane5 launch
system would then be feasible.  The telecommunications restriction
remains the most significant negative however, given that only one
ground station is likely to be affordable for the mission.

Nonetheless, the current orbit is expected to be a halo orbit, of
fairly high amplitude, around L2. This orbit avoids any eclipses
during the whole mission, and maximises the uniformity of illumination
of the spacecraft shielding by the Sun, Earth and Moon, thereby
minimising variable thermal gradients on the payload. The orbit also retains
maximum efficiency for illumination of the solar arrays, and naturally
provides an efficient scanning pattern around the sky. Since this
requires the spacecraft to be spinning around its line of sight towards
the ground station, efficient communications requires a set of
phased-array antennae distributed around the service module.

Even the most benign location will still mean that the spacecraft
requires occasional orbital correction, due to such factors as solar
wind variations. Since sudden accelerations are
very disadvantageous for astrometry, the expected means of spacecraft
control will involve Field Emission Electric Propulsion, FEEP. This
system delivers  thrust (a few mN) through acceleration of
cesium atoms. It provides very smooth accelerations over a mission
lifetime with extremely low-mass requirements. FEEP thrusters are
being space qualified late in 1998, and are expected to be available,
and suitable, for GAIA.

\section{CONCLUSION}

GAIA is a proposed global astrometric mission which builds on the
success of space astrometry demonstrated by HIPPARCOS. The substantial
gains in accuracy, sensitivity, and sample size now available,
relative to HIPPARCOS, allow extension of our scientific horizons to
one of the great intellectual challenges: understanding the origin and
evolution of our own galaxy, the Milky Way. In addition GAIA will
determine the distribution function of planetary systems, by
identifying and surveying the complete sample of the 300,000 nearest
stars, and significantly improve tests of standard general relativity.
The technological challenges in implementing the GAIA promise are
considerable, but achievable.

\vskip 2.0cm


\noindent 1. F. de Felice, M.G. Lattanzi, A. Vecchiato, \&
P.L. Vernacca, ``General Relativistic Satellite Astrometry I''
{\sl Astron. Astrophys.}  {\bf 332} 1132-1141, 1998

\noindent 2. G. Gilmore \& E. Hoeg
``Key Questions in Galactic Structure with Astrometric Answers'' in {\sl
Future Possibilities for Astrometry in Space; ESA SP-379 } 
M. Perryman \& F. van Leeuwen eds, pp 95-98,  ESA, ESTEC, 1995

\noindent 3. G. Gilmore \& M. Perryman, 
``Stellar and Galaxy Kinematics: the Future'' in {\sl
Proper Motions and Galactic Astronomy: Astron. Soc. Pacific Conf Ser, 127}, 
Roberta M. Humphreys  ed, pp 191-197, ASP, 1997.

\noindent 4. G. Gilmore \& R.F.G. Wyse, 
``The Multivariate Stellar Distribution Function'' in {\sl
The Galaxy}, G. Gilmore \& B. Carswell eds, pp 247-280,  
Reidel, Dordrecht, 1987
 
\noindent 5. G. Gilmore, R.F.G. Wyse, \& K. Kuijken 
``Kinematics, Chemistry and Structure of the Galaxy''
{\sl Ann. Rev. Astron. Astrophys.}  {\bf 27 } 555-627, 1989.
 
\noindent 6. M.G. Lattanzi, A Spagna, A Sozzetti \& S. Casertano,
`` GAIA and the Hunt for Extra-Solar Planets'', in {\sl
HIPPARCOS: Venice '97; ESA SP-402}, B. Battrick ed, pp 755-759,
ESA, ESTEC, 1997

\noindent 7. M.A.C. Perryman, L. Lindegren, \& C. Turon, 
``The Scientific Goals of the GAIA Mission'' in {\sl
HIPPARCOS: Venice '97; ESA SP-402}, B. Battrick ed, pp 743-748,
ESA, ESTEC, 1997

 \end{document}